\documentclass[12pt]{article}
\usepackage{epsfig}
\usepackage{amsmath}
\usepackage{amssymb}

\usepackage{epsfig,graphicx}
\newcommand{\bk}{\mbox{\boldmath$k$}}
\newcommand{\br}{\mbox{\boldmath$r$}}
\newcommand{\bp}{\mbox{\boldmath$p$}}
\newcommand{\beq}{\begin{equation}}
\newcommand{\eeq}{\end{equation}}

\begin{document}

\title{ On the possible mixing of the electron capture and the positron
emission channels in nuclear decay
}
\author{V.I. Isakov\thanks{E-mail: Vadim.Isakov@thd.pnpi.spb.ru}\\
\date{}
Petersburg Nuclear Physics Institute, 188300 Gatchina, Russia}
\maketitle
\vspace*{-0.5cm}
\begin{abstract}
On the basis of the idea of mixing (interaction) between the electron
capture and the positron emission channels in the $\beta^+$ decay in the
cases when both  channels are energetically allowed, we attempt
to explain  oscillations of the $K$-capture rates that were
possibly seen in the recent experiment.
\end{abstract}
\noindent PACS: 23.40.Bw; 23.40.-s

\vspace{0.5cm}
{\Large  1. Introduction}
\vspace{0.5cm}

In the papers \cite{Litvinov}, \cite{Kienle2} the authors observed
the time-dependent
oscillations with the period $T^{lab}_{osc} \sim 7$ s of the electron
capture rates in the allowed Gamow--Teller (GT) decays of $^{140}$Pr and
$^{142}$Pm. The  preliminary result for $^{122}$I \cite{Kienle2} shows
$T^{lab}_{osc} \sim 6$ s. The authors measured the decay events in a
sequence of measurements, each of them was performed with a single
one-electron ion. These papers were attended by the theoretical article
\cite{Ivanov4}, where the authors tried to explain the
effect in the framework of the scheme of the neutrino oscillations.
This idea became an object of a lively discussion.

Trying to explain the oscillations seen in the experiment \cite{Litvinov}
in the framework of the more-or-less standard nuclear physics, we turn our
attention to another scenario. In the experiment \cite{Litvinov} the authors
observed the transition rate with respect to the electron capture (EC)
only.  The cases of $\beta^+$\, decay were out of the ``window" of
observations.  However, the possible coupling of the two above-mentioned
channels, due to a weak interaction between them, may lead to the
oscillations of the EC rate. Below we consider this possibility
qualitatively.

\vspace{0.5cm}
{\Large 2. Phenomenological approach}
\vspace{0.5cm}

First, we remind briefly the standard picture
of the neutrino oscillations in the $\beta$\, decay. The neutrino born
in the $\beta$\, decay is the electron one, $\nu_e$. However, the state
$|\nu_e\rangle$ is not the eigenstate of the total mass operator,
thus it is not a stationary one, if there exists a mixing between the
electron neutrino $\nu_e$ and  muon neutrino $\nu_\mu$. In the
presence of such a mixing, the eigenstates are the $|\nu_1\rangle$ =
$|\psi_1\rangle$ and $|\nu_2\rangle$ = $|\psi_2\rangle$ ones, each of them
is being a combination of $|\nu_e\rangle$ = $|\varphi_1\rangle$ and
$|\nu_\mu\rangle$ = $|\varphi_2\rangle$, while the states $|\nu_e\rangle$ and
$|\nu_\mu\rangle$ are not the stationary ones.  This leads to the time
oscillations of the value $_0\langle\nu_e|\nu_e\rangle_t$  due
to the transitions $\nu_e\to\nu_\mu$ and inverse ones.

Thus, in the presence of coupling we have
\begin{eqnarray}
&& |\varphi_1\rangle\,=\,\cos\vartheta|\psi_1
\rangle+\sin\vartheta|\psi_2\rangle\,,
\nonumber\\
&& | \varphi_2\rangle\,=\,-\sin\vartheta|\psi_1
\rangle+\cos\vartheta|\psi_2\rangle\,,
\nonumber\\
&&
|\varphi\rangle\ =\ ||M||\cdot |\psi\rangle\,, \quad \quad
||M||=\left|\left|{\cos\vartheta\ \sin\vartheta \atop -\sin\vartheta\
\cos\vartheta}\right|\right|,
\end{eqnarray}
where $|\psi_1\rangle$ and $|\psi_2\rangle$ are the eigenstates with the
account for mixing.

The inverse transformation looks as follows:
\beq
|\psi\rangle\ =\ ||M^{-1}||\cdot|\varphi\rangle\,, \qquad
||M^{-1}||\ =\left|\left|{\cos\vartheta\ -\sin\vartheta \atop
\sin\vartheta\quad \cos\vartheta} \right|\right|.
\eeq
The amplitude $A_{ki}=\,_0\!\langle\varphi_i|\varphi_k\rangle_t$
of transformation of the state
$|\varphi_i\rangle_0$ into the state $|\varphi_k\rangle_t$ is
\beq
A_{ki}(t,\vartheta, \Delta)\ =\ \sum_{j}
M_{kj}(\vartheta)S_{jj}(t)M^{-1}_{ji}(\vartheta)\,,
\eeq
where
\beq
||S(t)||\ =\ \left|\left|\begin{array}{cc}
e^{-i\Delta t/\hbar} & 0\\
0 & e^{i\Delta t/\hbar} \end{array} \right|\right|
\eeq
is the diagonal time-evolution matrix of the stationary states (we have
omitted here the insufficient common phase). Here
$\Delta=(E_1-E_2)/2$, while $E_1$ and $E_2$ are the energies of
stationary states $|\psi_1\rangle$ and $|\psi_2\rangle$. Thus, the
matrix $||A||$ has the form
\beq
||A||\ =\ \left|\left|\begin{array}{cc}
\cos(\Delta t/\hbar)-i\sin(\Delta t/\hbar)\cos(2\vartheta) &
i\sin(\Delta t/\hbar)\sin(2\vartheta)\\
i\sin(\Delta t/\hbar)\sin(2\vartheta) & \cos(\Delta
t/\hbar)+i\sin(\Delta t/\hbar)\cos(2\vartheta) \end{array}
\right|\right|.
\eeq

So, we have
\begin{eqnarray}
&&|A_{11}|^2=|A_{22}|^2=1-\sin^2(2\vartheta)\sin^2\left(\frac{\Delta
t}\hbar\right),
\nonumber\\
&&|A_{12}|^2=|A_{21}|^2=\sin^2(2\vartheta)
\sin^2\left(\frac{\Delta t}\hbar\right),
\nonumber\\
&& |A_{11}|^2+|A_{12}|^2=|A_{22}|^2+|A_{21}|^2=1.
\end{eqnarray}
For $\nu_{e}-\nu_{\mu}$ oscillations the
last equation in (6) is nothing but the unitarity relation.

Here we come to the difference between the $\nu_e-\nu_\mu$ and
$\rm EC-\beta^+$ oscillations. Instead of $|\varphi_1\rangle = |\nu_e\rangle$
and $|\varphi_2\rangle = |\nu_{\mu}\rangle$ we have now $|\varphi_1\rangle =
|\rm EC\rangle$ and $|\varphi_2\rangle = |\beta^+\rangle$,
that correspond to the transitions $|Z,N\rangle+e^-(1s)\to|Z-1,N+1
\rangle+\nu_e$ and $|Z,N\rangle+e^-(1s)\to|Z-1,N+1\rangle
+e^-(1s) +e^+ +\nu_e$ (in the last case the $1s$\,-electron is a
spectator).
The coupling between the states $|{\rm EC}\rangle=|\varphi_1\rangle$  and
$|\beta^+\rangle=|\varphi_2\rangle$ leads to their mixing and to the
energy splitting of the corresponding eigenstates
$|\psi_1\rangle, |\psi_2\rangle$,
as well as to the time dependence of $_0\langle\rm EC|\rm EC\rangle_t$.
In the case of the $\nu_e - \nu_{\mu}$ oscillations at $t=0$, we have the
electron neutrino only, while the muon neutrino appears only as a result of
oscillations. In our case we have not only the depopulation of the
$|\rm EC\rangle$ channel due to the $|{\rm EC}\rangle\to
|\beta^+\rangle$ oscillations, but also the population of this channel
due to transformations $|\beta^+\rangle\to|\rm EC\rangle$.
The $|\beta^+\rangle$ states appear not only due to the
$|\rm EC\rangle\to|\beta^+\rangle$ oscillations, but are
supplementarily settled also  in  the $\beta^+$ decay.
The probabilities of the electron capture and $\beta^+$ decay
are different, therefore  oscillations in both
channels arise. In this way, we obtain the formulae for the
transition rates in both channels:
\begin{eqnarray}
w_{\rm EC}(t) = w^0_{\rm EC}\cdot|A_{11}|^2 +
w^0_{\beta^+}\cdot|A_{12}|^2 =
\nonumber\\
 = w^0_{\rm EC}\left[1 + B\,\sin^2\left(\frac{\Delta t}{\hbar}\right)
\right]\,, B=  \frac{w^0_{\beta^+}-w^0_{\rm EC}}{w^0_{\rm EC}}\,
\sin^2(2\vartheta)\,;
\nonumber\\
w_{\beta^+}(t) = w^0_{\beta^+}\cdot|A_{22}|^2 +
w^0_{\rm EC}\cdot|A_{21}|^2 =
\nonumber\\
 = w^0_{\beta^+}\left[1 + D\,\sin^2\left(\frac{\Delta
t}{\hbar}\right) \right]\,, D=  \frac{w^0_{\rm
EC}-w^0_{\beta^+}}{w^0_{\beta^+}} \sin^2(2\vartheta)\,.
\end{eqnarray}
where $w^0_{\rm EC}$ and $w^0_{\beta^+}$ are the transition rates
for the electron capture as well as for $\beta^+$ decay in the absence
of mixing.

For the allowed Gamow--Teller transition we have
\begin{eqnarray}
w^0_{\rm EC}&=&\frac{m^5_ec^4}{2\pi^3\hbar^7}G^2_A\,B({\rm GT};
J_i \to J_f)\cdot2\pi^2\sum_{i}
\Psi_{i}^2(0)\left(\frac{E_{\nu_i}}{m_ec^2}\right)^2 ,
\nonumber\\
w^0_{\beta^+} &=& \frac{m^5_ec^4}{2\pi^3\hbar^7}G^2_A\,B({\rm GT};
J_i \to J_f)\cdot f(E_\beta,Z).
\end{eqnarray}

In (8) $G_A$ is the effective axial vector constant in nuclei
(see details in \cite{Isakov}),
$B(GT;J_i \to J_f)$ is the reduced transition probability for the
Gamow--Teller operator, $f(E_\beta,Z)$ is the integrated Fermi function
for the allowed $\beta$\, decay, $E_\beta$ is the maximal kinetic energy
of the positron
in the transformation $|Z,N\rangle \to |Z-1,N+1\rangle$, $E_{\nu}$
is the neutrino energy, while the
densities of the $K$-electrons at zero $\Psi_{i}^2(0)$ ($i=1,2, ...$)
that contribute into the $K$-capture rate for the allowed
transitions
are in  $(\hbar/m_ec)^{-3}$ units.
For the one-electron ion
$i=1$ only in (8). We see from Eq.~(7) that $w_{\rm
EC}(t)+w_{\beta^+}(t)=w^0_{\rm EC}+w^0_{\beta^+}=\lambda$, where
$\lambda$ is the decay constant in the exponential law
$e^{-\lambda t}$, while the counting rates are
\beq
\frac{dn_{\rm EC}}{dt}= w_{\rm EC}(t)\,N_0\,e^{-\lambda t}, \qquad
\frac{dn_{\beta^+}}{dt}=w_{\beta^+}(t)\,N_0\,e^{-\lambda t}, \qquad
\frac{d(n_{\rm EC}+n_{\beta^+})}{dt}=\lambda \,N_0\,e^{-\lambda t}.\qquad
\eeq

The equality $w_{EC}(t) + w_{\beta^+}(t) = \lambda$
is the unitarity relation in our case. We see that the total
transition rate (in both channels) does not depend on time, thus
we again have the exponential law for the decay of the parent nucleus.
Taking the values of nuclear masses from \cite{iaea} and
using the beta-decay Tables \cite{Dzelepov} for the values of
$f(E_{\beta},Z)$ and $\Phi_{i}^2(0)$ (for the one-electron atom we took
a half of the density of the $K$-shell in the neutral atom
as the single-particle functions of the 1s electron are in practice the
same in the one-electron and neutral atoms
), we obtain for the decay of the one-electron $^{142}$Pm  $w^0_{\rm
EC}/w^0_{\beta^+} \approx 0.12$. Thus,
\begin{eqnarray}
w_{\rm EC}(t) &=& w^0_{\rm EC}\left[1+7.33\sin^2(2\vartheta)
\sin^2\left(\frac{\Delta t}\hbar\right)\right],
\nonumber\\
w_{\beta^+}(t) &=& w^0_{\beta^+}\left[1-0.88\sin^2(2\vartheta)
\sin^2\left(\frac{\Delta t}\hbar\right)\right].
\end{eqnarray}
Note that $7.33\,w^0_{\rm EC}=0.88\,w^0_{\beta^+}$. We see from (10) that
the situation in $^{142}$Pm is favorable for oscillations in the EC
channel due to its small partial width.

For the decay of the one-electron $^{140}$Pr we obtain
$w^0_{EC}/w^0_{\beta^+} \approx 0.41$. So
\begin{eqnarray}
w_{\rm EC}(t) &=& w^0_{\rm EC}\left[1+1.44\sin^2(2\vartheta)
\sin^2\left(\frac{\Delta t}\hbar\right)\right],
\nonumber\\
w_{\beta^+}(t) &=& w^0_{\beta^+}\left[1-0.59\sin^2(2\vartheta)
\sin^2\left(\frac{\Delta t}\hbar\right)\right].
\end{eqnarray}
From the analysis of the experimental data \cite{Litvinov} for $^{142}$Pm
and $^{140}$Pr one
may easily conclude on the values of the parameters $\Delta$
and $\vartheta$ entering Eqs.~(10) and (11).  The pre-exponential
factor in the counting rates was defined in \cite{Litvinov} as
$\left[1 + A\,{\rm cos}(\omega_{osc}t + \phi)\right]$, where $A =
0.23(4)$, $T^{lab}_{osc}=2\pi/\omega_{osc}=7.10(22)$\,s for $^{142}$Pm
and $A=0.18(3)$, $T^{lab}_{osc}=7.06(8)$\,s  for $^{140}$Pr. It is
better to work in the system where the $^{142}$Pm and $^{140}$Pr
nuclei are at rest. Here $T_{osc} \approx 7/\gamma \approx 5$ s, where
$\gamma =1.43$ is the corresponding Lorentz factor \cite{Kienle2}.
The phase $\phi$ was not defined in \cite{Litvinov} as the
experimental data are absent at small values of $t$.  Our approach
leads to $\phi=\pi\,, A= B/(2+B) \approx B/2$ and $\omega_{osc}=2\Delta/
\hbar$ (see Eq.~(7)),
while by using Eqs.~(10), (11) and the values of
$A$ shown above we obtain $\vartheta \approx 8^\circ,\, \Delta = \pm
0.407(13)\cdot 10^{-15}$ eV for $^{142}$Pm and $\vartheta \approx
16^\circ,\, \Delta = \pm 0.410(5)\cdot 10^{-15}$ eV for $^{140}$Pr.
The patterns of oscillations based on the above-mentioned discourse are
shown in Fig.~1 for the decay of $^{142}$Pm and in Fig.~2 for the decay
of $^{140}$Pr.  Here  in both cases $w^0_{\beta^+}$ are larger than
$w^0_{\rm EC}$. As a result, in the presence of oscillations the
transition rates for the electron capture are  higher than those in the
absence of oscillations.  At the same time, the situation is opposite
for the $\beta^+$ decays.  One can easily consider the corresponding
integral effect. Let us introduce the total numbers of decays in the
corresponding  channels as
\beq
N({\rm EC})= N_0 \int\limits^\infty_0
w_{\rm EC}(t)\,e^{-\lambda t}dt\,, \qquad N(\beta^+)=  N_0
\int\limits^\infty_0 w_{\beta^+}(t)\,e^{-\lambda t}dt\,.
\eeq
Then, we can easily obtain
\beq \frac{N(\rm EC)}{N(\beta^+)} = \frac{w^0_{\rm
EC}}{w^0_{\beta^+}} \cdot \frac{\left[1+\frac{w^0_{\beta^+}-w^0_{\rm
EC}}{2\,w^0_{\rm EC}}\, \sin^2(2\vartheta)\, \frac{1}{ 1 +
(\frac{\lambda\,\hbar}{2\,\Delta})^2 }\right]} {\left[1+\frac{w^0_{\rm
EC} -w^0_{\beta^+}}{2\,w^0_{\beta^+}}\, \sin^2(2\vartheta)\,
\frac{1}{1+(\frac{\lambda\,\hbar}{2\,\Delta})^2 }\right]}\,.
\eeq

In  cases of interest, when one can observe several oscillations
at the time interval $\tau=1/\lambda$ we have $\Delta/\hbar \gg \lambda$
($\lambda=0.017$ s$^{-1}$ and $\Delta/\hbar =0.67$ s$^{-1}$
for  $^{142}$Pm). Then we have

\beq
\frac{N(\rm EC)}{N(\beta^+)} = \frac{w^0_{\rm EC}}{w^0_{\beta^+}}\cdot
\frac{\left[1+\frac{w^0_{\beta^+}-w^0_{\rm EC}}{2\,w^0_{\rm EC}}\,
\sin^2(2\vartheta)\right]}
{\left[1+\frac{w^0_{\rm EC} -w^0_{\beta^+}}{2\,w^0_{\beta^+}}\,
\sin^2(2\vartheta)\right]} \, .
\eeq

Formula (14) can be easily obtained if we substitute
$\sin^2(\Delta t/ \hbar)$ in (7) by its average value of 1/2, the
averaging is over the time interval $\delta t$ more than $\hbar/\Delta$.
This is in some sense equivalent to  averaging over the ensemble  of
initial nuclei, that are formed during the time interval greater than
$\hbar/\Delta$, as the time counter is switched on for each nucleus
in the moment of its formation.

For the decay of the one-electron $^{140}$Pr we have
$N(\rm EC)/N(\beta^+)$=1.34$\cdot w^0_{\rm EC}/w^0_{\beta^+}$.
For the neutral atom of $^{140}$Pr the ratio $w^0_{\rm EC}/
w^0_{\beta^{+}}$ should be twice as much as for the one-electron ion,
i.e. it should be equal 0.82. If we consider the electron capture from
higher $s$-orbits, this value by using \cite{Dzelepov}  is 0.97.
For neutral $^{140}$Pr
the values of $w^{0}_{\beta^{+}}$ and $w^{0}_{\rm EC}$ are
close to each other, and we can see from the Eq.(14) that
$N(\rm EC)/N(\beta^{+})\approx$ $w^{0}_{\rm EC}/w^{0}_{\beta^{+}}\cdot
1.06$. The experimental data on the ratio of the $K$-capture to
the $\beta^{+}$ decay of the neutral $^{140}$Pr are rather
vague. In refs. \cite{Brabek} -- \cite{Campbell}, they are 0.897,
0.74, 0.9 and 0.85 correspondingly, giving the average value and the
standard  deviation equal to 0.846(75). This value should be compared to
the theoretical value $N({\rm EC}\,\, (1s))/N(\beta^{+}) = 0.82 \cdot 1.06
=0.87$. Thus, the accuracy of  experimental data
is insufficient to make definite conclusion on the enhancement
of the EC($1s$) rate as compared to the standard calculations, which do not
include the mixing of  two decay channels. For $^{142}$Pm with two
electrons on the $K$ shell (or for the neutral atom), formula (14) shows
the increase of $N(\rm EC)/N(\beta^{+})$ to be
about 16 $\%$ as compared to the standard calculations (i.e.
$w^{0}_{\rm EC}/w^{0}_{\beta^{+}} \approx 0.257$, see \cite{Tuli}).
At the same time, the experimental value of this ratio is equal to
0.297(45) \cite{Firestone}.
So one can see the increase of the ratio $N(\rm EC)/N(\beta^+)$
as compared to the theoretical value $w^0_{\rm EC}/w^0_{\beta^+}$ obtained
without mixing of final states, the difference is in
accordance with our prediction, though the experimental errors are
large. The previously mentioned estimations used the values of
$\vartheta$ from the one-electron ions.

The  systematics of the ratios
$(w^{0}_{\rm EC}(1s)/w^{0}_{\beta^{+}})_{exp}/
(w^{0}_{\rm EC }(1s)/w^{0}_{\beta^{+}})_{th}$ shown in Fig.~3 demonstrates
that they often differ from the unity up to 10$\%$, the deviations
are in both sides. Our discourse leads to a  small tendency for the
above-mentioned ratios to be a bit smaller than the unity at
$(w^0_{\rm EC}/w^0_{\beta^+})_{th} > 1$, and to be a bit larger
than the unity at $(w^0_{\rm EC}/w^0_{\beta^+})_{th} < 1$.

\vspace{0.5cm}
{\Large 3. Microscopical evaluation}
\vspace{0.5cm}

Here we try to explain the above-discussed picture by using certain
qualitative arguments. Suppose that there exists some additional
interaction $H_{\rm w}$, which couples the EC and $\beta^+$ channels.
First, we determine the
magnitudes of the corresponding matrix elements using the values of
$\Delta$ and $\vartheta$ shown above. These  matrix
elements may be easily determined from  secular
equation obtained in the two-level scheme with the $\rm|EC\rangle$
and $|\beta^+\rangle$ as basic functions. Including the interaction
$H_{\rm w}$, we introduce the quantities
$E_\beta=V_{\beta^+,\beta^+}=\langle \beta^+|H_{\rm w}|\beta^+\rangle$,
$E_{\rm EC}=V_{\rm EC,\rm EC}=\langle{\rm EC}|H_{\rm w}|{\rm EC}\rangle$,
as well as $V_{\rm EC,\beta^+}=\langle{\rm EC}|H_{\rm w}|\beta^+
\rangle$. Then one can easily obtain
\beq
\delta\ \equiv\ \frac{\langle{\rm EC}|H_{\rm w}|{\rm EC}\rangle-\langle
\beta^+|H_{\rm w}|\beta^+\rangle}2\ = \Delta\cos(2\vartheta),
\, \,\,\Delta=(E_1-E_2)/2\,\, ;
\eeq
At the same time,
\beq
V_{\rm EC,\beta^+}\ =\ \langle{\rm EC}|H_{\rm w}|\beta^+\rangle= \Delta
\sin(2\vartheta)\,\,\,,
\eeq
$V_{\rm EC,\beta^+}= 0.11\cdot10^{-15} \rm eV$ for $^{142}$Pm and
$V_{\rm EC,\beta^+}= 0.22\cdot10^{-15} \rm eV$ for $^{140}$Pr.

We see from (15) and (16) that corresponding matrix
elements are very small, of the order of $10^{-16}$eV. They may arise
due to the weak interaction stipulated by both the neutral and
charged weak currents. The  matrix element $V_{\rm EC,\beta^+}$ is
graphically shown in Fig.~4, while the $V_{\beta^+,\beta^+}$ is
represented in Fig.~5. At the same time, one can put the value of
$V_{\rm EC,EC}$ to be equal to zero, because  only the higher-order
diagrams contribute here.

First, consider the matrix element $V_{\rm EC,\beta^+}$ that is shown
in Fig.~4, replacing the positrons by electrons with inverse momenta.
In this case we have the $\nu - e$ interaction with the  matrix element
that accounts for both  $Z$ and  $W$ bosons and looks as follows
\cite{Okun}\,:
\begin{eqnarray}
M &=& \frac{G}{\sqrt2}\langle\bar{u}_{e2}|g_L\gamma_{\alpha}
\left(1+\gamma_5\right) +g_R\gamma_{\alpha}\left(1-\gamma_5\right)|
u_{e1}\rangle\,
\langle\bar{u}_{\nu 2}|\gamma^{\alpha}\left(1+\gamma_5
\right)|u_{\nu 1}\rangle \, \approx
\nonumber\\
&\approx&
\frac{G}{\sqrt{2}}\langle\bar{u}_{e2}|\gamma_{\alpha}+\frac{1}{2}
\gamma_{\alpha}\gamma_5|u_{e1}\rangle\, \langle\bar{u}_{\nu 2}|
\gamma^{\alpha}\left(1+\gamma_5\right)|u_{\nu 1}\rangle\,.
\end{eqnarray}

In (17), we have $g_{L}=\frac{1}{2} + \sin^2\Theta_W, g_R=
\sin^2\Theta_W$. Here $\Theta_W$is the Weinberg angle $(\sin^2\Theta_W
\approx \frac{1}{4})$, while $G = G_V/\cos\Theta_C$, where $\Theta_C$ is
the Cabibbo angle $(\cos\Theta_C = 0.974)$. From the experiments on the
investigation of the superallowed $\beta$-transitions between the
isoanalog states of nuclei, it follows \cite{Hardy} that the weak vector
coupling constant $G_V=1.395\cdot10^{-49}$\,erg$\cdot$cm$^3$ = 87.08
eV$\cdot$fm$^3$. Below we consider the contact type of the interaction
between the weak currents that enter formula (17), use the plane
waves for the unbounded leptons and take the wave function of the $1s$
electron (this function includes the angular part $Y_{00}$) in the
non-relativistic form
\beq \varphi_{1s}(\br)\ =\
\frac1{\sqrt\pi}\,a^{3/2}e^{-ar}\quad  (a= \frac
{Zm_ee^2}{\hbar^2}=\alpha Z/\,^-\hspace{-0.55em}\lambda_e
=1.154\cdot10^{-3}\,\rm fm^{-1}\ for\ ^{142}Pm)\,,
\eeq
where $^-\hspace{-0.55em}\lambda_{e}$ is the Compton wavelength of the electron.
The calculations were also performed
under the assumption of the uniform angular distributions of the
entering fast leptons.  In addition, we took into account that the
process of the $\beta$ decay is mediated by left currents and the
energies of positrons  are rather high. Thus, we considered
the positrons as having the right spirality. We also considered
that the average momentum of the bounded $1s$ electron is much less
than $(m_e\,c)$, and we neglected the corresponding contributions.
Then we can schematically represent the $V_{\rm EC,\beta^+}$ as

\begin{eqnarray}
V_{\rm EC,\beta^+}&\sim& 2G\,\,\frac{8\,\sqrt\pi}{V^{3/2}_{eff}}
\frac{a^4}{(k^2+a^2)^2}\cdot\frac1{a^{3/2}}\bigg[1-(k^2+a^2)^{3/2}
\frac{r_{\max}}{2ak}e^{-ar_{\max}}\sin(kr_{\max}+\gamma)
\nonumber\\
&&+\quad
\frac{(k^2+a^2)}{2ak}\,e^{-ar_{\max}}\sin(kr_{\max}+\sigma)\bigg],
\end{eqnarray}
where
$$
\gamma=\arctan\left(\frac ka\right), \qquad
\sigma=\arctan\left(\frac{2ak}{a^2-k^2}\right).
$$

In (19), $V_{eff}$ is the effective volume for the leptons in the
continuum, $V_{eff}\approx\frac43\pi r^3_{\max}$, where $r_{\max}$ is
of the order of some units of $1/a$, $r_{\max} \sim C/a$,
while $\bk=\frac1{\hbar}\,\bp$,
where $\bp=\bp(\nu_e,{\rm EC})-\bp(\nu_e,\beta^+)-\bp(e^+,\beta^+)$.

Note that really the spectra of both  positron and  neutrino,
which are produced in the $\beta^+$ decay, are not monochromatic.
Averaging the values of $k$ over the corresponding distributions,
as it was in the calculations of the integrated $\beta$ decay
Fermi function $f(E,Z)$ and supposing the uniform angular
distributions of the unbound leptons, we obtain for $^{142}$Pm
the value $\bar p\sim\sqrt{\bar {p^2}}\sim5.5$\,MeV/$c$, which
corresponds to $\bar k\!\sim\!\rm3.0\!\cdot\!10^{-2}\,fm^{-1}$.

We mention that the averaging of the second and  third
terms in the right-hand side of (19) leads to their vanishing. In any
case, their contribution may be neglected by the absolute value as
compared to the contribution of the first term at $r_{\max}\geq 8/a$.
So, the value of $C \sim 8$ defines the upper limit of integration, and
in this way the volume of the interaction (the normalization
volume for the unbound leptons that is of the order of the volume of
the neutral atom).

Now we come to the evaluation of the matrix element
$V_{\beta^+,\beta^+}$. The corresponding
diagrams are shown in Fig.~5. The diagram (a) ($e - \nu$
interaction) is represented by the formula (17), while the $e -
e$ interaction, that is mediated by the $Z$ boson only, is shown in
the diagram (b). It is described by the matrix element

\begin{eqnarray}
M &=& \frac{G}{\sqrt{2}}\langle\bar{u}_{e2}|g_L\gamma_{\alpha}
\left(1+\gamma_5 \right)+g_R\gamma_{\alpha}\left(1-\gamma_5\right)
|u_{e1}\rangle \times
\\
&& \times\
\langle\bar{u}_{e4}|g_L\gamma^{\alpha}\left(1+\gamma_5\right)+g_R
\gamma^{\alpha}\left(1-\gamma_5\right)|u_{e3}\rangle \, \approx
\frac{G}{\sqrt{2}}\,\frac{1}{4}\,\langle\bar{u}_{e4}|\gamma_{\alpha}\gamma_5
|u_{e1}\rangle\,\langle\bar{u}_{e4}|\gamma^{\alpha}\gamma_5|u_{e3}\rangle\,,
\nonumber
\end{eqnarray}

as here $g_L=-\frac{1}{2}+\sin^2\Theta_W$ and $g_R=\sin^2\Theta_W$.

The formula for the matrix elements $V_{\beta^+,\beta^+}$ of the
interaction corresponding to the diagrams shown in  Fig.~5 looks as
follows:

\begin{eqnarray}
V_{\beta^+\beta^+} &\sim& G\,\,\frac{16}{V_{eff}}
\frac{a^4}{(k^2+4a^2)^2}\bigg[1-(k^2+4a^2)^{3/2}\frac{r_{\max}}{4ak}\,
e^{-2ar_{\max}}\sin(kr_{\max}+\gamma_1)
\nonumber\\
&& +\quad
\frac{(k^2+4a^2)}{4ak}\,e^{-2ar_{\max}}\sin(kr_{\max}+\sigma_1)\bigg],
\end{eqnarray}
where
$$
\gamma_1=\arctan\left(\frac k{2a}\right), \qquad \sigma_1=\arctan\left(
\frac{4ak}{4a^2-k^2}\right).
$$


As a result, we obtain for $^{142}$Pm
$V_{\rm EC,\beta^+} \approx 0.056\cdot 10^{-15}$ eV and
$V_{\beta^+,\beta^+} \approx  1.4\cdot 10^{-15}$ eV $(\delta \approx
 0.7\cdot 10^{-15})$ eV, which may be compared  with the results of
Eqs. (15) and (16), $V_{\rm EC,\beta^+} = 0.11 \cdot 10^{-15}$ eV and
$V_{\beta^+,\beta^+} \approx 0.8 \cdot 10^{-15}$ eV.
Note that the mixing angle $\vartheta$ for $^{140}$Pr
($Q_{\rm EC} \approx$ 3.4 MeV) is larger than
for $^{142}$Pm ($Q_{\rm EC} \approx$ 4.8 MeV). This fact finds an evident
explanation if we look at formula (19), where $V_{{\rm EC},\beta^{+}}
\sim 1/(a^2 + {\bar k}^2)^2$ ($\vartheta$ is approximately proportional
to $V_{{\rm EC},\beta^{+}}$ at small $\vartheta$, while $\bar k$ is
larger for $^{142}$Pm, than for $^{140}$Pr). At the same time, it is
difficult to understand, why the periods of oscillations $T_{osc}$ are
very close to each other in the cases of decay of $^{142}$Pm and
$^{140}$Pr.

\vspace{0.5cm}
{\Large  4. Hyperfine interaction}
\vspace{0.5cm}

Here we evaluate the energy splitting of levels due to  magnetic
fields that exist in the accelerator as a possible source of oscillations.
We also evaluate the effects of the hyperfine interaction.

The magnetic moments of the ground-state $1^+$ levels of $^{140}$Pr and
$^{142}$Pm are unknown by now. However, it follows from the
single-particle scheme that these odd--odd nuclei have the
configuration $\{p\, 2d_{5/2},n\, 2d_{3/2}; I^\pi=1^+\}$. The average
value of the magnetic moment of the proton on the orbit $\{p\,
2d_{5/2}\}$ obtained from the experimental data on the proton-odd
nuclei $^{141}$Pr and $^{143}$Pm is $\sim4.0\,\mu_N$
$(\mu_N=e_{p}\hbar/2m_Nc)$, while the average value of the magnetic
moment of the neutron on the orbit $\{n 2d_{3/2}\}$ is
$\sim1.0\,\mu_N$; this value is determined from the neutron-odd nuclei
$^{139}$Ce and $^{141}$Nd. The above-mentioned evaluations used the
fact that the magnetic moment of the state $|j^{n_{odd}}\, s=1\,
;J=j\,\rangle$ for the lowest seniority $``s"$ does not depend
on $n_{odd}$.  In the case of the two-particle configuration
$|j_1j_2I\rangle$ we have the following formula for the gyromagnetic
ratio of this state:
\beq
g_I\ =\ \frac{g_1+g_2}2 +
\frac{g_1-g_2}2\cdot\frac{j_1(j_1+1)-j_2(j_2+1)}{I(I+1)}\,,
\eeq
where $g_1$ and $g_2$ are the gyromagnetic ratios for the states
$|j_1\rangle$ and $|j_2\rangle$. In this way we have
$\mu_I(I^\pi=1^+;\,
^{140}$Pr)$\,\approx\mu_I(I^\pi=1^+;\,
^{142}$Pm)$\,\approx\!2.3\,\mu_N$.  For the the
magnetic field $H\sim1\,\, T$, we obtain the magnitude of the
interaction of nuclear magnetic moment with the external field equal to
$\sim\mu_I\cdot H=0.7\cdot10^{-7}$eV. At the same time, the
interaction of the electron spin with the magnetic field is much
stronger, $\sim \mu_B\cdot H =0.6\cdot10^{-4}$eV
$(\mu_B=e\hbar/2\,m_ec)$. However, there exists also the interaction
between the electron and the nucleus spins. For an electron
on the $1s$ orbit we have
\beq
E_{I,s;F}\ =\ \frac83\,\mu_B\cdot\mu_I\cdot\frac{F(F+1)-I(I+1)-s(s+1)}I
\cdot a^3\,,
\eeq
where $I\!=\!1$ and $s\!=\!1/2$ are spins of the nucleus and  electron
respectively, while $F\!=\!I\pm1/2$ is the total spin of the
one-electron ion.  The magnitude of (23) is equal to
$0.3\cdot[F(F+1)-11/4]$ eV that is much larger than the interactions of
magnetic moments of the electron and of the nucleus with the magnetic
field. Thus, the two spins are strictly coupled to each other. By using
formula (22) we find for $F\!=\!3/2$\, the value of
$\mu_{F=3/2}=\mu_I(I\!=\!1^+)+\mu_B\!\approx\!\mu_B$ while for
$F\!=\!1/2$ we obtain
$\mu_{J=1/2}=\frac23\mu_I(I\!=\!1^+)-\frac13\mu_B\!\approx\!-\frac13\mu_B$.
So the corresponding energy splitting due to the magnetic field is of
the order of $10^{-4}\,$eV, which is much greater than  characteristic
magnitude of the effect $(\sim 10^{-16})\,$eV seen in the experiment
\cite{Litvinov}. However, we mention here the
paper \cite{Pavl}, where it was shown that under certain conditions one
may expect modulation of the EC decay rate due to the resonance
multiphoton transitions between the magnetic substates of the ground
$F=1/2$ state of $^{140}$Pr$^{58+}$, or $^{142}$Pm$^{60+}$.

Another source of splitting in the one-electron ion may be the weak
interaction in the neutral channel between the electron and the nucleus.
Neglecting the spin structure of this interaction, we obtain the
evaluation of its strength, being equal by the order of magnitude to
$\Psi_{1s}^2(0)\cdot G_{V} \,\,\sim \,\,10^{-7}$\,eV, that is also much
greater, than the value of $\Delta$, seen in the experiment.

We mention here that, to our opinion, the $\nu_e - \nu_{\mu}$ and
$\nu_{\mu} - \nu_{\tau}$ oscillations, that explain the experiments showing
the suppression of the Solar neutrino and  reactor antineutrino,
as well as the atmospheric muon neutrino fluxes, do not refer to the
results of \cite{Litvinov}. The above-mentioned experiments correspond
to the mass differences $\Delta(m^2)_{e,\mu} \sim 10^{-4}$ eV$^2$
($\Delta m_{e,\mu} \sim 10^{-2}$ eV) and $\Delta m_{\mu,\tau} \sim
10^{-1}$ eV,
these numbers are larger in 14 -- 15 orders of magnitude  than the
value of $\Delta$ observed in \cite{Litvinov}.

\vspace{0.5cm}
{\Large  5. Two-electron atoms}
\vspace{0.5cm}

Here, we consider the difference between the $K$-capture rates in the
one-electron and the two-electron ions of $^{142}$Pm and $^{140}$Pr.
For the one-electron ions we have the transition between the initial
state $|I_i=1, s_e=1/2; F_i\rangle$ and the final state $|I_f=0,
s_\nu=1/2; F_f\rangle$. As the Hamilton operator conserves the total
angular momentum, we have $F_i=F_f=1/2$. For the Gamow-Teller
transition in the $\beta^+$ and the EC channels we have
\beq
\hat
H_{int} = G_A\left(\sum_{i}\mbox{\boldmath$\sigma$}_L(i)\tau_L^+
(i)\right)\cdot\left(\sum_{k}\mbox{\boldmath$\sigma$}_N(k)\tau_N^-(k)
\right)\,,
\eeq
where the summations over ``$i$" and ``$k$" refer to electrons and
nucleons, correspondingly, while $\tau^{\pm}$ are the operators that
change the charge of a particle by one.
By using the standard Racah algebra \cite{Shalit} we obtain for
the transition matrix element $M$
\beq
M=\langle I_f=0, s_\nu=1/2;
1/2|\hat H_{int}|I_i=1, s_e=1/2; 1/2\rangle\,=
\eeq
$$=\,
G_{A}\,\,W(1/2,1/2,1,\,1\,;0\,,1)\langle
1/2||\mbox{\boldmath$\sigma$}||1/2 \rangle \, \langle
I_f=0||(\sum_{k}\mbox{\boldmath$\sigma$}
_N(k)\tau_N^-(k))||I_i=1\rangle $$
$$=\, -G_A\langle I_f=0\|\hat m(GT)\|I_i=1\rangle\, \psi_{1s}(0)\,,$$
where
$\psi_{1s}(0)$ is the upper component of the
single-particle electron wave function at
zero. Thus,
\beq |M|^2=G_A^2\psi_{1s}^2(0)\langle I_f=0\|\hat
m(GT)\|I_i=1\rangle ^2 \equiv 3\cdot
G_A^2\psi_{1s}^2(0)B_{GT}(1^+\rightarrow 0^+)\,,
\eeq
as
\beq
B_{GT}(I_i\rightarrow I_f)=\frac{1}{2I_i+1}\langle I_f=0\|\hat
m(GT)\|I_i=1\rangle ^2\,.
\eeq

At the same, the initial state may have the value of the total spin $F$
equal to both 1/2 and 3/2, while only the transition from the $F=1/2$
really happens. Thus, we should multiply the value $|M|^2$ which defines
the transition rate and is given by
(26) by the factor 1/3. This factor was considered in Eq.~(8), where
$\Psi_i^2(0) = \psi_{1s}^2(0)=\frac{1}{4\pi}\,|g_{1s}(0)|^2\,$.

In the case of the two-electron atom we have the  state $|I_i=1,
(s_e=1/2)^2J=0; F=1\rangle$ as the initial one, while the final state
is $|I_f=0,(s_e=1/2, s_\nu=1/2)J=1; F=1\rangle$. By considering the
lepton system we should obligatory account for the antisymmetrization
between the remaining electron and the neutrino, as we have the process
where these leptons transform into each other. As a result, we have
\beq |M|^2=2\cdot G_A^2\,\psi_{1s}^2(0)\, B_{GT}(1^+\rightarrow 0^+)\,.
\eeq
In this way, we obtain formula (8) for the two-electron atom, where
$\Psi_{i}^2(0)=2\,\psi_{1s}^2(0)=2\,\frac{1}{4\pi}\,|g_{1s}(0)|^2\,$.

By considering the process of possible time oscillations of the
$K$-capture rate in the neutral atoms (here we consider the
two-electron atoms) one should also take into account the many-body
effects and the Pauli principle.
These effects can reveal themselves both in variation of the energy
shifts and in variation of the mixing amplitude. Here, the two-electron
wave function $|(1s)^2J=0\rangle$ looks as
\beq
|(1s_{1/2})^2J=0\rangle
=\varphi_{1s}(\mbox{\boldmath$r$}_1)
 \varphi_{1s}(\mbox{\boldmath$r$}_2)\frac{\chi_{1/2}(1)
\chi_{-1/2}(2)-\chi_{1/2}(2)\chi_{-1/2}(1)}{\sqrt{2}}\,.
\eeq
If we consider the energy shift in the $\beta^+-$channel and average
over the directions of the electron and the neutrino, then the diagonal
matrix element of the interaction increases by two as compared to the
case of one-electron ion shown in Fig.~5, i.e. the value of $\Delta$
increases by two. At the same time, the matrix element of mixing
becomes equal to $\frac{1}{\sqrt{2}}\, (V_{{\rm EC},\beta^{+}}(m_{1s}=1/2)-
V_{{\rm EC},\beta^{+}}(m_{1s}=-1/2))$,
where $V_{{\rm EC},\beta^{+}}(m_{1s})$ are the matrix
elements shown in the Fig. 4. As we do not have the selected axis
and average over the directions of the particles, both
these matrix elements are equal to each other. As a result, the mixing
between the $\beta^+$ and the EC channels is absent $(\vartheta=0)$, and
thus the oscillations disappear. If we adopt this assertion, we conclude
that all filled $(ns)$ shells do not contribute into the oscillation
effect. The electron structures for the neutral atoms of $^{142}$Pm and
$^{140}$Pr are the $(4f_{5/2})^5 (6s_{1/2})^2$ and
$(4f_{5/2})^3(6s_{1/2})^2$ ones correspondingly (we show only the
electrons
above the Xe core). Thus, only the electrons with $\ell\neq 0$ can
contribute, their possible contribution is negligibly small.
We mention here the paper \cite{Mikhailov} where it was shown that that
the spectra of the bound-state $\gamma$-quanta following the radiative
electron capture are different in the cases of one and two-electron
ions, this difference is also due to the Pauli principle.

Here we indicate the analogy of the $\beta^+/$EC decays with the decays
of $K$-mesons. In the last case,
due to the second order weak interaction that
does not conserve the strangeness $S$, the real eigenstates are not the
$|K^0\rangle$ and $|\bar K^0\rangle$, but
the $|K_1^0\rangle =(|K^0\rangle+|\bar K^0\rangle)/\sqrt{2}$ and
$|K_2^0\rangle =(|K^0\rangle -|\bar K^0\rangle )/\sqrt{2}$ ones.
At the same time, if we
neglect the CP-violation, the $|K^{0}_2\rangle $ meson, due to the
structure of its wave function does not decay via the $2\pi$ mode (only
via the $3\pi$ one), and is a long-lived particle,
$|K^0_2\rangle = |K_L\rangle$, while
$|K^0_1\rangle = |K_S\rangle$ is a short-lived one. Thus,
in the case of $K$-mesons the $|K^0_2\rangle$-meson is
the long-lived one, while in our case the mixing between the two
channels is close to zero due to the Pauli principle.  The states
$|K^0\rangle$
and $|\bar K^0\rangle$ oscillate in time, while the total decay rate
(in both channels) is the sum of the two exponents \cite{Lipkin}.
It is appropriate here to notice the difference with the oscillations
of $K$ mesons. In the last case the mixing between the $|K_0 \rangle$
and $|\bar K_0 \rangle$ mesons is the maximal one, $\vartheta = \pi
/4$, and the energy shift between the $|K^0_1 \rangle$ and $|K^0_2
\rangle$ is due only to the non-diagonal mixing. The experimental data
show, that in our case the mixing angle between the EC and the $\beta^+$
channels is 3 -- 6 times less. Thus, we are obliged to introduce the very
small additional energy shift, $V_{\beta^+\beta^+}$. In this regard we
have the situation intermediate between the oscillations of $K$-mesons
and the neutrino oscillations.

Here we mention the experimental paper \cite{Vetter}, performed with the
ensemble of neutral atoms of $^{142}$Pm arising as
a result of the reaction
$^{124}$Sn ($^{23}$Na, 5$n$)$^{142}$Pm in a sequence of short
irradiation bursts. The duration of each burst was much less
than the period of the expected oscillations, while the interval
between the bursts was much more than the half-life of the initial
$1^+$ state of $^{142}$Pm.  The best fit corresponds to amplitude
$A=0.0145(74)$, while $T_{osc}=3.18$\, s (in the system where the
$^{142}$Pm nucleus is at rest; here the results \cite{Litvinov} for
the one-electron ions  are $A =0.23$ and $T_{osc}=5$\, s).
Thus, we performed model calculations that correspond to the duration
of the irradiation burst equal to 0.5 s, as it was in \cite{Vetter}.
The counting rate $N(t, \tau)$, see Eq.~(30) is normalized in such a way,
that $N(t\to 0, \tau \to 0) \to 1$.

\begin{eqnarray}
&&\hspace*{-0.5cm}
N(t, \tau)=\frac{1}{(1-A)\,\tau}\int_{-\tau}^{\, 0}
\exp(-\lambda(t-t'))\cdot[1+A\, \cos(\omega(t-t')+\pi)]dt'=
\frac{\exp(-\lambda t)}{(1-A)\tau}
\nonumber\\
&&\times\  \left[\frac{1-\exp(-\lambda
\tau)}{\lambda} +\frac{A}{\sqrt {\omega^2+\lambda^2}}
[\cos(\omega t +\pi + \psi)-\cos(\omega(t+\tau)+\pi+\psi)\exp(-\lambda
\tau)]\right];
\nonumber \\
&& \hspace*{2cm} \psi\ =\ \arctan(\omega\,/\,\lambda)\,.
\end{eqnarray}

The pattern of oscillations of the $K$-capture rate by $^{142}$Pm is
shown in the Fig.~6, for different values of the entering parameters.
For the one-electron ion we have the same diagram, as in Fig.~1, i.e.
the interval $\tau = 0.5$ \,s is small, and the oscillation picture
is not washed away.
If we adopt the values of $\Delta$ and $\vartheta$ in the
two-electron ion the same as for the one-electron ion, the amplitude of
oscillations  $A$ attenuates due to the decrease of the factor
$\frac{w_{\beta^+}^0- w_{\rm EC}^0}{w_{\rm EC}^0}$, see Eq.~(7), while
the frequency of oscillations remains the same, $\omega_2=\omega_1$. If
we increase the value of $\Delta$ by two $(\Delta_2=2\Delta_1)$ by
remaining the value of the coupling matrix element the same as in the
one-electron ion, the frequency of oscillations also increases by two
$(\omega_2=2\omega_1)$, while their amplitude $A$ further decreases
(as the mixing angle $\vartheta\sim V_{{\rm EC},\beta^{+}}/\Delta$
decreases). If, under $\Delta_2=2\Delta_1$, we decrease the coupling
matrix element, we have $\omega_2=2\,\omega_1$, while
the amplitude of oscillations $A$ decreases still more, and we approach
to the exponential decay law.

One can see that the  pattern of oscillations of the $K$-capture rate in
neutral $^{142}$Pm shown by us before is in a qualitative agreement with
the result of \cite{Vetter}.

\vspace{0.5cm}
{\Large  6. Conclusion}
\vspace{0.5cm}

In this paper, in the framework of the hypothesis of mixing between the
electron capture and  $\beta^{+}$ channels we tried to explain
the oscillations of the $K$-electron capture rate that were presumably
seen
in the recent experiments. Such a mixing leads to a small variation of
the ${\rm EC}/\beta^{+}$ ratio in the decay of the one-electron ions
and to an even much smaller variation of this ratio for the ensemble of
neutral atoms as compared to standard calculations. The precision of
both the available experimental data as well as of the theoretical
calculations  of this ratio is not sufficient to make  conclusions on
this subject.

According to our hypothesis, the time oscillations
of the electron capture rate should be strongly hindered
if one makes an experiment analogous to \cite{Litvinov} but with
the two-electron ions, or neutral atoms of $^{140}$Pr, $^{142}$Pm,
or $^{122}$I. This statement is confirmed by the results of
\cite{Vetter}. The most direct way to check the hypothesis
is to observe the time-antiphase oscillations in the $\beta^{+}$-decay
branch in the decay of one-electron $^{140}$Pr, where one can expect
the amplitude of oscillations $A$ of about 0.08, this amplitude is
0.03 in $^{142}$Pm, as the effect of the $\beta^+$ oscillations
increases by the decrease of $w(\beta^+)/w(\rm EC)$, i.e. by the
decrease of $Q_{\beta}$. The preliminary experimental data
\cite{Kienle2} relating to the $\beta^+$ branch in the decay of the
one-electron ion of $^{142}$Pm give the result $A = 0.03(3)$.

Our approach for calculation of the oscillation parameters is rather
simplified, especially the introduction of the effective interaction
volume for the unbounded leptons. Actually, unbound
leptons may leave the atom before the interaction. One
important remark is in place here. It was noted by
\cite{Merle} that the Darmstadt effect can not arise due
to the interaction in the final state. Our approach is not the
study of the final state interaction. Really, the effect arises
due to the interference of the two possible paths of evolution:
the direct $K$-capture, and the population of
the $K$-capture channel through the intermediate $\beta^+$-decay state.
As a result, the quantum beatings arise. As the mixing matrix element
is very small, the period of these beatings is very large. The more
detailed analysis of this effect should be the subject of a separate
investigation.

This paper was the subject of numerous discussions with my colleagues,
particulary relating to the mechanism of a possible mixing.
However, the time scale of the effect, if it really exists, denotes
the weak interaction between the objects of large dimension being the
only source of the necessary energy splitting.

The author is grateful to Ya.I.~Azimov, B.~Fogelberg, F.F.~Karpeshin,
Yu.N.~Novikov, M.G.~Ryskin, V.R.~Shaginuan, and M.B.~Trzhaskovskaya for
discussions and useful critical comments.

This work was performed under the support of the Russian
Foundation for Basic Research (grant No RSGSS-3628.2008.2).


\newpage
\begin{figure}[h]
\vspace{-2.0cm}
\centerline{\epsfxsize=14.0cm\epsfbox{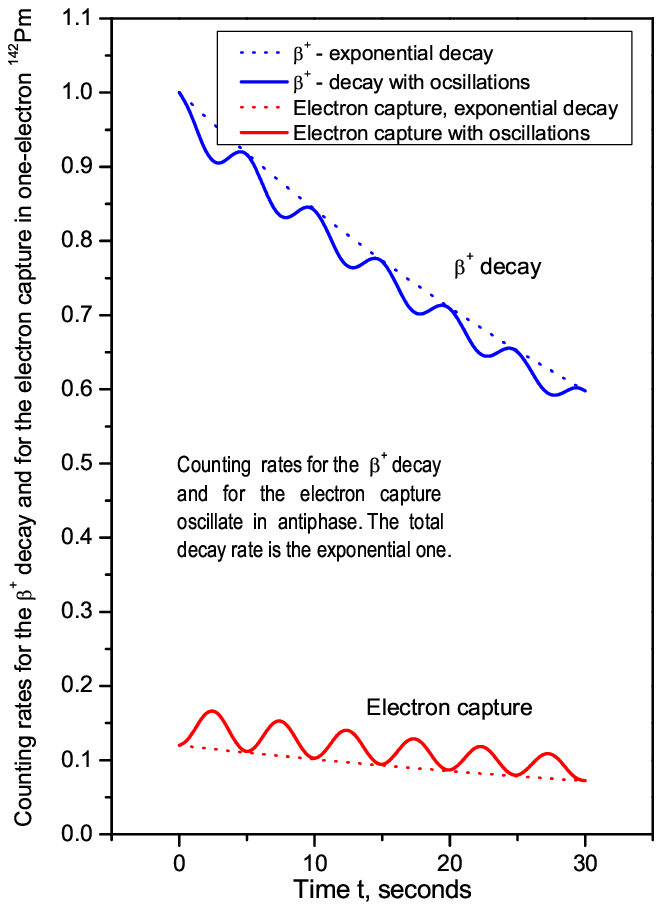}}
\vspace{1.0cm}
{\bf Fig. 1} Counting rates for the electron
capture and the $\beta^+$ decay for the one-electron ion
$^{142}$Pm in the presence of the weak coupling between the two
decay channels; $T_{1/2} = 40.5\,$s, $T_{osc} = 4.96\,$s (in the system,
where the $^{142}$Pm$^{+60}$ ions are at rest). The
counting rate in the $\beta^{+}$ channel at $t=0$ is adopted to be
unity.
\end{figure}

\newpage
\begin{figure}[h]
\vspace{-2.0cm}
\centerline{\epsfxsize=14.0cm\epsfbox{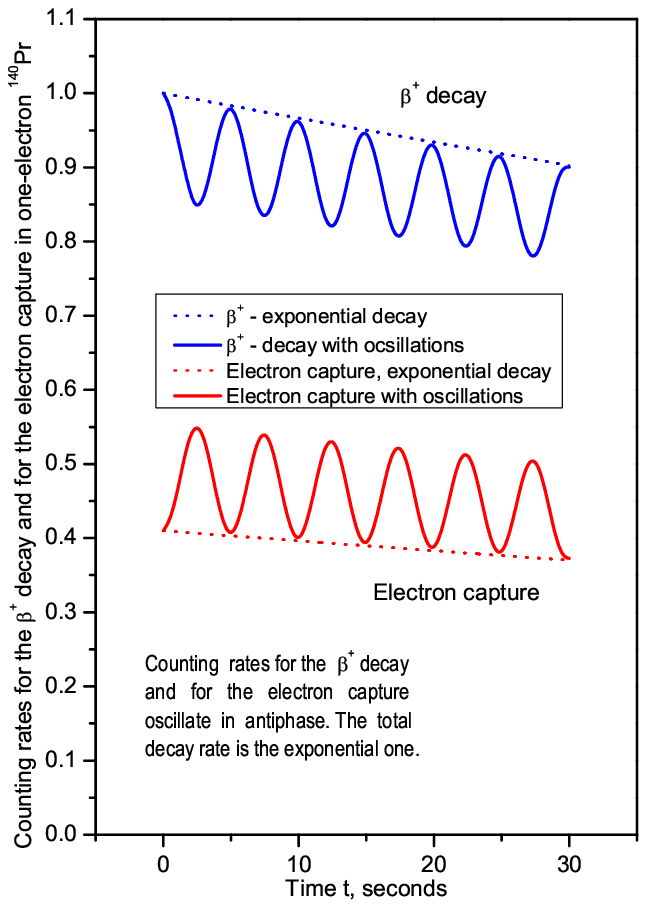}}
\vspace{1.0cm}
{\bf Fig. 2} Counting rates for the electron capture and  $\beta^+$
decay for the one-electron ion  $^{140}$Pr in the presence of the
weak coupling between the two decay channels; $T_{1/2} = 3.39\,$min,
$T_{osc} = 4.94\,$s (in the system, where the $^{140}$Pr$^{+58}$ ions
are at rest). The counting rate in the $\beta^{+}$ channel at
$t=0$ is adopted to be unity.
\end{figure}

\newpage
\begin{figure}[h]
\vspace{2.0cm}
\centerline{\epsfxsize=17.0cm\epsfbox{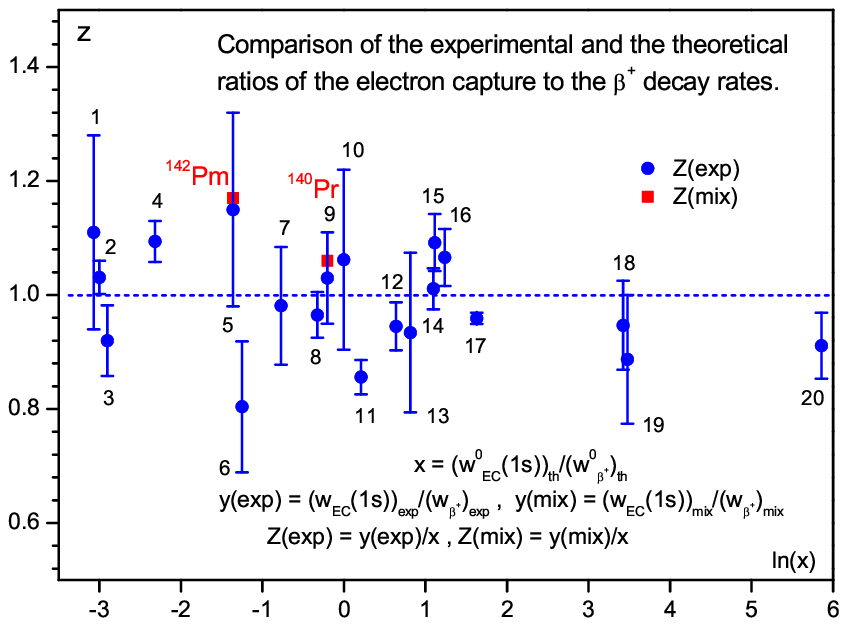}}
\vspace{0.0cm}
{\bf Fig. 3} Experimental data versus theoretical ratios of the
$w^0_{\rm EC}(1s)/w^0_{\beta^+}$ in neutral atoms as a function of
$(w^0_{\rm EC}(1s))_{th}/(w^0_{\beta^+})_{th}$. Only the allowed
Gamow--Teller transitions are shown here. The ratio Z(mix) is
calculated by using the mixing parameters from the one-electron ions.
The decrease of the mixing angle $\vartheta$ in the two-electron or
neutral atoms leads to the tendency Z(mix)$\to 1\,$.
The notations are as follows: \\
{\bf 1}: $^{140}$Eu $(1^+ \to 0^+, 1.51\,$s);
{\bf 2}: $^{44}$Sc $(2^+ \to 2^+, 3.07\,$h);
{\bf 3}: $^{91}$Mo $(9/2^+ \to 9/2^+, 15.49\,$min);\\
{\bf 4}: $^{22}$Na $(3^+ \to 2^+, 2.60\,$y);
{\bf 5}  $^{142}$Pm $(1^+ \to 0^+, 40.5\,$s);
{\bf 6}: $^{61}$Cu $(3/2^- \to 3/2^-, 3.33\,$h);\\
{\bf 7}: $^{134}$La $(1^+ \to 0^+, 6.45\,$min);
{\bf 8}: $^{48}$V $(4^+ \to 4^+, 15.97\,$d);
{\bf 9}: $^{140}$Pr $(1^+ \to 0^+, 3.39\,$min);\\
{\bf 10}: $^{143}$Sm $(3/2^+ \to 5/2^+, 8.75\,$min);
{\bf 11}: $^{120}$Sb $(1^+ \to 0^+, 15.89\,$min); \\
{\bf 12}: $^{52}$Mn $(6^+ \to 6^+, 5.59\,$d);
{\bf 13}: $^{64}$Cu $(1^+ \to 0^+, 12.70\,$h);
{\bf 14}: $^{89}$Zr $(9/2^+ \to 9/2^+, 78.41\,$h);
{\bf 15}: $^{89}$Zr $(1/2^- \to 3/2^-, 4.16\,$min);
{\bf 16}: $^{116}$Sb $(8^- \to 7^-, 60.30\,$min);  \\
{\bf 17}: $^{58}$Co $(2^+ \to 2^+, 70.86\,$d);
{\bf 18}: $^{65}$Zn $(5/2^- \to 3/2^-, 244.06\,$d); \\
{\bf 19}: $^{141}$Nd $(3/2^+ \to 5/2^+, 6.45\,$min);
{\bf 20}: $^{107}$Cd $(5/2^+ \to 7/2^+, 6.50\,$h).
\end{figure}

\newpage

\begin{figure}[h]
\vspace{-1.0cm}
\centerline{\epsfxsize=14.0cm\epsfbox{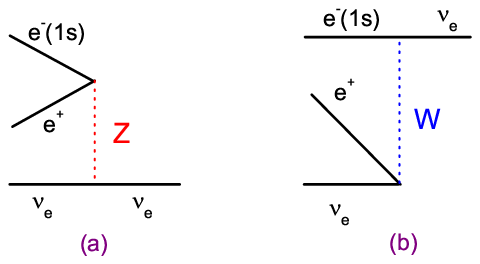}}
\vspace{0.0cm}
{\bf Fig. 4} The diagrams demonstrating  possible
coupling between the electron capture and  $\beta^+$ channels.
\end{figure}

\begin{figure}[h]
\vspace{-1.0cm}
\centerline{\epsfxsize=14.0cm\epsfbox{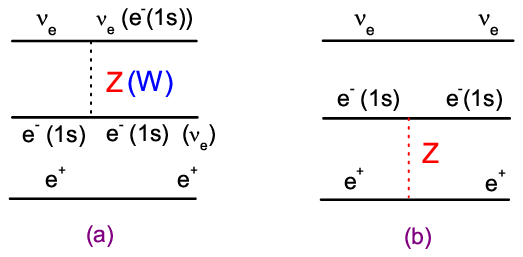}}
\vspace{0.0cm}
{\bf Fig. 5} The diagram showing the energy shift in the $\beta^+$
channel. The exchange by the $\gamma$-quantum in  the
diagram (b) is not considered, as the corresponding effect is included in
the Coulomb functions of the charged leptons.
\end{figure}

\newpage
\begin{figure}[h]
\vspace{-2.0cm}
\centerline{\epsfxsize=14.0cm\epsfbox{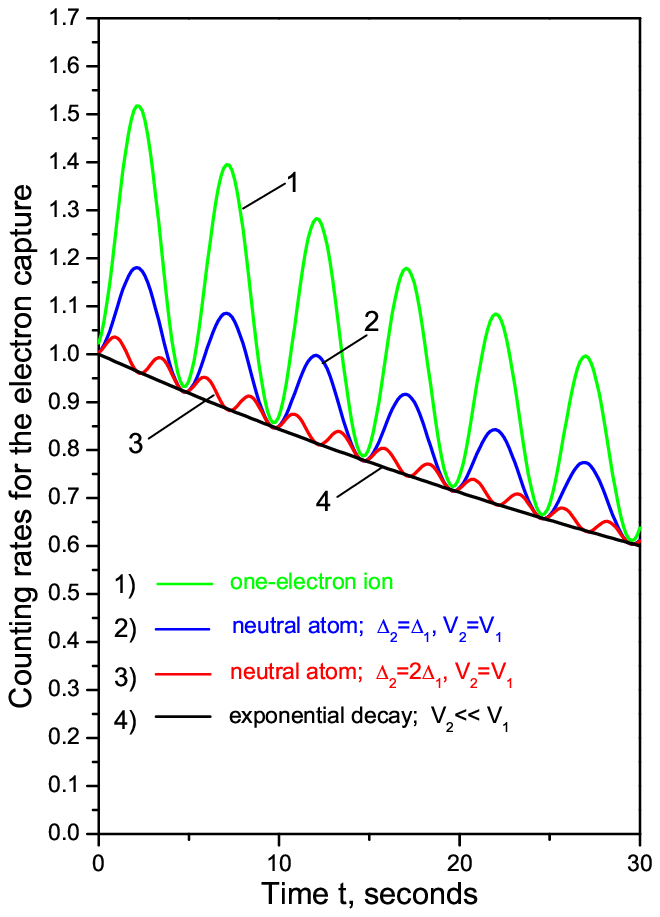}}
\vspace{0.0cm}
{\bf Fig. 6} Model calculation of the decay law relative to the electron
capture for the ensemble of one-electron ions or the neutral atoms
of $^{142}$Pm as a function of the entering parameters. The $^{142}$Pm
nuclei are supposed to be produced in the irradiation bursts with duration
$\tau = 0.5$\, s.  The time reading begins just after the termination
of the burst.
\end{figure}

\end{document}